# Digital Ecosystem for FAIR Time Series Data Management in Environmental System Science

J. Bumberger[1,2,3], M. Abbrent[1,2], N. Brinckmann[4], J. Hemmen[1,2], R. Kunkel[5], C. Lorenz[6], P. Lünenschloß[1,2], B. Palm[1,2], T. Schnicke[2,7], C. Schulz[2,7], H. van der Schaaf[8], and D. Schäfer[1,2]

[1] Helmholtz Centre for Environmental Research – UFZ, Department Monitoring and Exploration Technologies, Permoserstraße 15, 04318 Leipzig, Germany
[2] Helmholtz Centre for Environmental Research – UFZ, Research Data Management - RDM, Permoserstraße 15, 04318 Leipzig, Germany
[3] German Centre for Integrative Biodiversity Research (iDiv) Halle-Jena-Leipzig, Puschstraße 4, D-04103 Leipzig, Germany
[4] Helmholtz Centre Potsdam – GFZ German Research Centre for Geoscience, Department Geoinformation, eScience Centre, Telegrafenberg, 14473 Potsdam, Germany
[5] Forschungszentrum Jülich - FZJ, Institute of Bio- and Geosciences (IBG), Agrosphere (IBG-3), Wilhelm-Johnen-Straße, 52428 Jülich, Germany
[6] Karlsruhe Institute of Technology – KIT, Institute of Meteorology and Climate Research Atmospheric Environmental Research (IMK-IFU), Kreuzeckbahnstraße 19, 82467 Garmisch-Partenkirchen, Germany
[7] Helmholtz Centre for Environmental Research – UFZ, IT Department, Permoserstraße 15, 04318 Leipzig, Germany
[8] Fraunhofer Institute of Optronics, System Technologies and Image Exploitation IOSB, Fraunhoferstraße 1, 76131 Karlsruhe, Germany

**Abstract**
Addressing the challenges posed by climate change, biodiversity loss, and environmental pollution requires comprehensive monitoring and effective data management strategies that are applicable across various scales in environmental system science. This paper introduces a versatile and transferable digital ecosystem for managing time series data, designed to adhere to the FAIR principles (Findable, Accessible, Interoperable, and Reusable). The system is highly adaptable, cloud-ready, and suitable for deployment in a wide range of settings, from small-scale projects to large-scale monitoring initiatives. The ecosystem comprises three core components: the Sensor Management System (SMS) for detailed metadata registration and management; time.IO, a platform for efficient time series data storage, transfer, and real-time visualization; and the System for Automated Quality Control (SaQC), which ensures data integrity through real-time analysis and quality assurance. The modular architecture, combined with standardized protocols and interfaces, ensures that the ecosystem can be easily transferred and deployed across different environments and institutions. This approach enhances data accessibility for a broad spectrum of stakeholders, including researchers, policymakers, and the public, while fostering collaboration and advancing scientific research in environmental monitoring.

**Metadata**

| Nr | Code metadata description | *Please fill in this column* |
|---|---|---|
| C1 | Current code version | **SMS:** 1.16.1 <br> **time.IO:** 0.1 <br> **SaQC:** 2.6 |
| C2 | Permanent link to code/repository used for this code version | **SMS:** https://hdl.handle.net/20.500.14372/SMS-Repository <br> **time.IO:** https://codebase.helmholtz.cloud/ufz-tsm <br> **SaQC:** https://git.ufz.de/rdm-software/saqc |
| C3 | Permanent link to reproducible capsule | **SMS:** https://zenodo.org/doi/10.5281/zenodo.13329925 <br> **time.IO:** https://zenodo.org/doi/10.5281/zenodo.8354839 <br> **SaQC:** https://zenodo.org/doi/10.5281/zenodo.5888547 |
| C4 | Legal code license | **SMS:** EUPL-1.2 <br> **time.IO:** EUPL 1.2 <br> **SaQC:** GNU GPL 3.0 |
| C5 | Code versioning system used | GIT |
| C6 | Software code languages, tools and services used | **SMS:** Docker, Elasticsearch, MinIO, nginx, PostGIS Python, TypeScript <br> **time.IO:** Alpine, CAdvisor, Django, django-helmholtz-aai, Docker CE, Docker Compose, FastAPI, FROST, Grafana, MinIO, Mosquitto MQTT Broker, nginx, NumPy, Pandas, Python Click, TimescaleDB, Tomcat <br> **SaQC:** Python |
| C7 | Compilation requirements, operating environments and dependencies | **SMS:** Docker, Docker Compose <br> **time.IO:** Docker, Docker Compose <br> **SaQC:** Python 3.8+ |
| C8 | If available, link to developer documentation/manual | **SMS:** https://hdl.handle.net/20.500.14372/SMS-Readme https://hdl.handle.net/20.500.14372/SMS-Wiki <br> **time.IO:** https://codebase.helmholtz.cloud/ufz-tsm <br> **SaQC:** https://rdm-software.pages.ufz.de/saqc/index.html |

| C9 | Support email for questions | **SMS:** sms-core-team@listserv.dfn.de |
| --- | --- | --- |
| | | **time.IO:** rdm-contact@ufz.de |
| | | **SaQC:** saqc-support@ufz.de |

## 1. Motivation and significance

Climate change, biodiversity loss, environmental pollution and related anthropogenic impacts are leading to significant pressures on the world's ecosystems and their functions, requiring a comprehensive quantification of these impacts (Stocker et al. 2012). To address this, large-scale and standarised monitoring observatories such as NEON (Loescher et al. 2017), eLTER (Mollenhauer et al. 2018; Ohnemus et al. 2024) and TERENO (Zacharias et al. 2011; Zacharias et al., 2024) have been established to provide essential data for the long-term monitoring with sensor systems and modelling of environmental systems. In addition, e.g. MOSES (Weber et al. 2023) investigates the evolution and impacts of highly dynamic, often extreme events (e.g., heat waves or hydrological extremes) using a systemic monitoring approach. These event-oriented, cross-compartment datasets are needed to understand the impacts of climate change, biodiversity loss and pollution, and to develop effective adaptation strategies. These observation networks are continuously expanding in terms of sensor density and geographical coverage. However, the resulting increase in data volumes poses challenges in handling and processing these data streams in real time while adhering to FAIR principles (Findable, Accessible, Interoperable, Reusable; Wilkinson et al. 2016) and standardised interfaces, as well as associated metadata. Sensor management, effective storage and automatic quality assurance of time series data require the development of scalable, high-performance, and transferable data infrastructures (Mons et al. 2017; Koedel et al. 2022). The integration of sensor data into such data infrastructures ensures subsequent real-time availability for further analysis, advanced data science methods, quantification of remote sensing based indicators (e.g. Selsam et al. 2024), and integration into models and information systems. Furthermore, robust time series data infrastructures facilitate the integration of data from different sources into distributed data infrastructures at institutional, regional, national and continental levels, thereby the dissemination of data to a wide range of stakeholders, including scientists, policy makers, resource managers and the general public. Historically, such data management infrastructures have been developed independently within institutions and integrated into existing IT frameworks. The rapid growth of sensors and observatories has outstripped the scalability of these systems. Consequently, interdependencies have arisen that make comprehensive overtaking resource-intensive, often resulting in incremental improvements rather than holistic redevelopment. In response, we have pursued an innovative approach to collaboratively design and systematically develop a comprehensive data infrastructure specifically for time series data management with the following characteristics and requirements. These are:

- **Interoperability:** Use of standardised interfaces and metadata standards as well as standard protocols for sensor integration
- **Modularity:** Application of Sensor management components with persistent identifiers, sensor data infrastructure component with the ability to connect different storage solutions, and quality assurance/data processing component, each independently deployable
- **Transferability and cloud readiness:** Use open source solutions in a microservice architecture, and container-based deployment for easy scalability and integration with existing IT infrastructures
- **Authentication system:** Leverage cross-institutional identity management to enable collaboration and the ability to use own authentication systems
- **User-friendly:** With simple responsive web interfaces for operation and integrated data viewers for data dissemination
- **Generic application:** Applicable to all domains using sensor-based data, to go beyond earth system science

## 2. Digital Ecosystem for FAIR Time Series Data Management

The conception of the digital ecosystem for FAIR time series data management began in 2019, building on the experience gained in the development of an interoperable data infrastructure for the TERENO observatories since 2009 (Zacharias et al. 2011). Particular attention has been paid to the modularity of the overall system and to the ability of the components to operate independently. One of the key innovations of this integrated system is its transferability and usability by other research institutions, authorities, companies and organizations worldwide. It has been designed to be easily deployable and adaptable to different needs, significantly improving the accessibility and usability of time series data for end users, operators, system integrators, and maintainers. The digital ecosystem for FAIR time series data management has been designed under specific conditions and requirements and consists of three core components or independently usable systems:

(i) **Sensor Management System (SMS):** Facilitates the detailed registration of sensors with an internal and globally available persistent identifier (EUDAT B2INST[1]) and the management of sensor metadata using international standards (OGC SensorML[2]). Authentication is managed by the Helmholtz AAI[3], which allows all users from institutions connected to GÉANT EduGAIN[4] to log in. It also includes the option to use an alternative institutional or other identity provider system.

(ii) **Data Infrastructure (time.IO):** Provides the infrastructure for storing and managing time series data. Standard input protocols ((S-)FTP and MQTT[5]) are used for data ingestion, while the standardised OGC SensorThings API (OGC STA[6]) is used for data access. Authentication is handled by the Helmholtz AAI[3], enabling all users from institutions connected to GÉANT EduGAIN[4] to log in. Additionally, it offers the option to use an alternative institutional or other identity provider system.

(iii) **System for Automated Quality Control (SaQC):** Enables real-time analysis, annotation, and processing of data using predefined or custom quality schemes in real-time, including end-to-end metadata enrichment for each data point. This can be done using pre-configured standard procedures and user-configured methods for a variety of environmental variables.

---

[1] https://b2inst.gwdg.de/
[2] https://www.ogc.org/standard/sensorml/
[3] https://hifis.net/doc/helmholtz-aai/
[4] https://edugain.org/
[5] https://mqtt.org/
[6] https://www.ogc.org/standard/sensorthings/

The integration of the three main components is achieved through appropriate interfaces between the systems, enabling the establishment of a comprehensive digital ecosystem for sensor-based measurements (Figure 1).

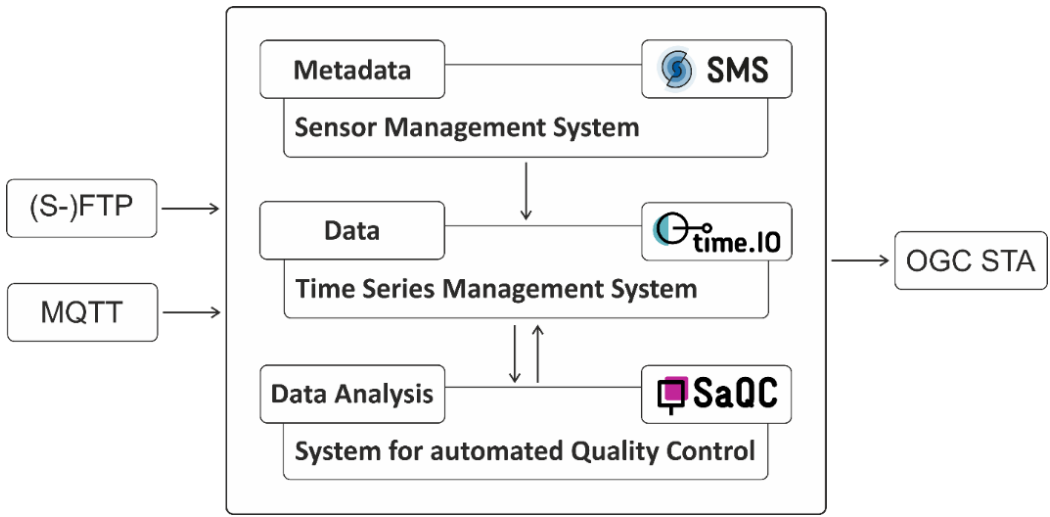

*Figure 1: Software architecture of the Digital Ecosystem for FAIR Time Series Data Management with the three main components for the management of sensor metadata (SMS), data (time.IO), and analysis or automated quality control of the data (SaQC).*

The overall system includes all essential components, such as user-centric web-based front-ends for the sub-components, a versatile data integration layer, robust time-series database, efficient object storage, real-time quality control, and comprehensive real-time data visualization functions. It supports modern and legacy data transfer protocols ((S-)FTP and MQTT) and ensures compliance with OGC standards for data access and sensor metadata. In addition, the fully integrated containerized solution offers the convenience of rapid deployment and seamless integration with existing institutional services such as databases, identity providers, and object stores. The following sections will provide detailed descriptions of the three main components of the Digital Ecosystem for FAIR Time Series Data Management.

## 2.1 Sensor Management System (SMS)

The Sensor Management System (SMS) is the foundation for sensor metadata handling. It allows detailed registration and management of sensors and specific measurement parameters, documenting changes over time. Following the JSON:API[7] specification and adhering to international standards, such as OGC SensorML, SMS ensures that data sources are discoverable and accessible through standardized interfaces. This is particularly important for integration into larger data infrastructures and for providing consistent metadata management. It uses a container-based deployment model for easy integration and scalability within existing IT infrastructures.

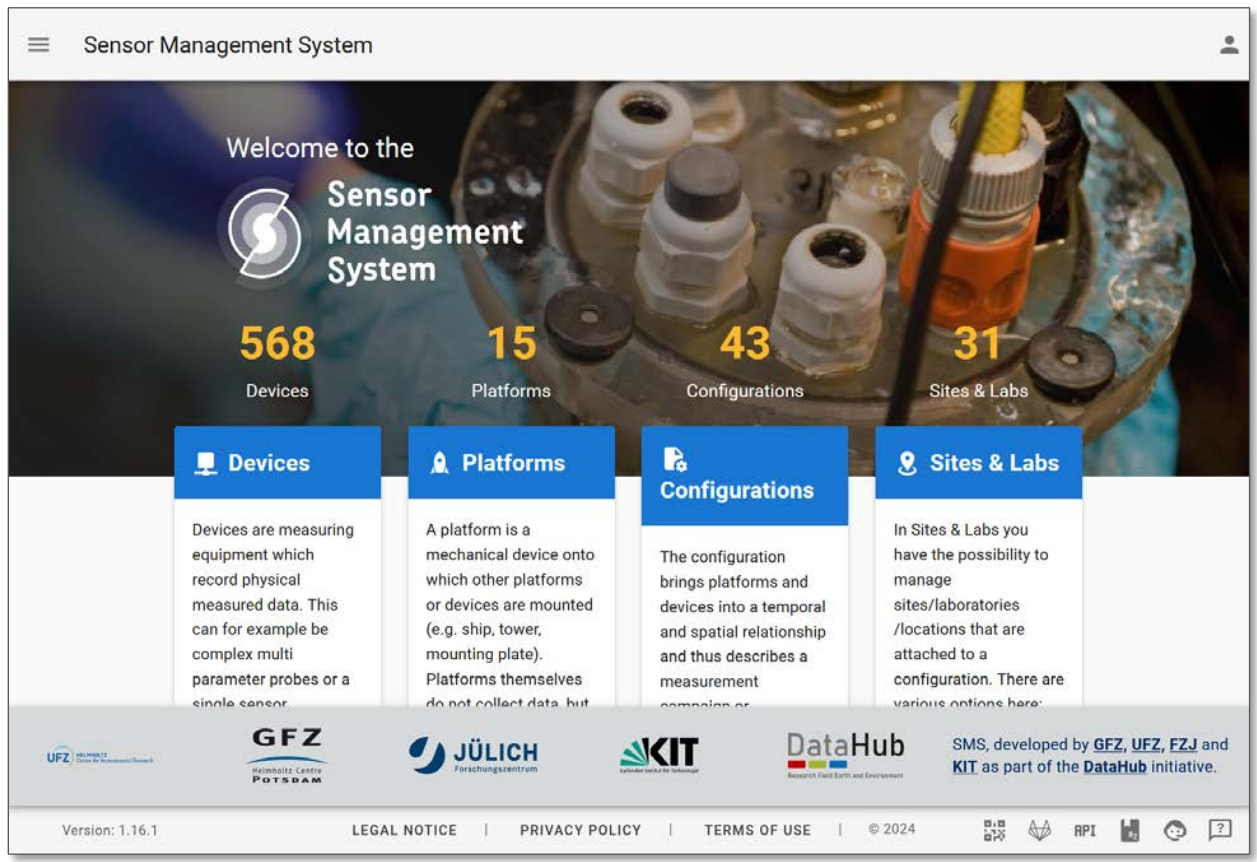

*Figure 2: Front-End SMS. The Sensor Management System (SMS) is designed to handle the acquisition and management of metadata from various sensors.*

---

[7] https://jsonapi.org/

Its main features include:

- **Sensor Registration and management**: Allows for the detailed registration and management of sensors and specific measurement parameters, documenting changes over time. The permanent registration of the instruments is done with the globally available persistent identifier (EUDAT B2INST)
- **Standard interfaces**: Uses JSON:API and follows international standards (OGC SensorML) for the provision of sensor metadata and measurement parameters.
- **Accessibility**: Ensures that all sensors with their metadata are findable and accessible through standardized interfaces.
- **Deployment**: Provides a container-based deployment model for easy integration and scalability within existing IT infrastructures.

For more details, see: Brinckmann et al. 2024 and/or
https://hdl.handle.net/20.500.14372/SMS-Wiki

## 2.2 time.IO

Building on the foundation of sensor metadata using SMS, time.IO provides the infrastructure for storing and managing time series data. It supports the entire lifecycle of time series data, providing efficient data transfer and storage, real-time data visualisation using Grafana, and integrated data analysis and quality control with SaQC. The container-based deployment model facilitates easy integration and scalability within existing IT infrastructures. time.IO also links to the SMS for consistent and standardised metadata management, ensuring a cohesive data management process.

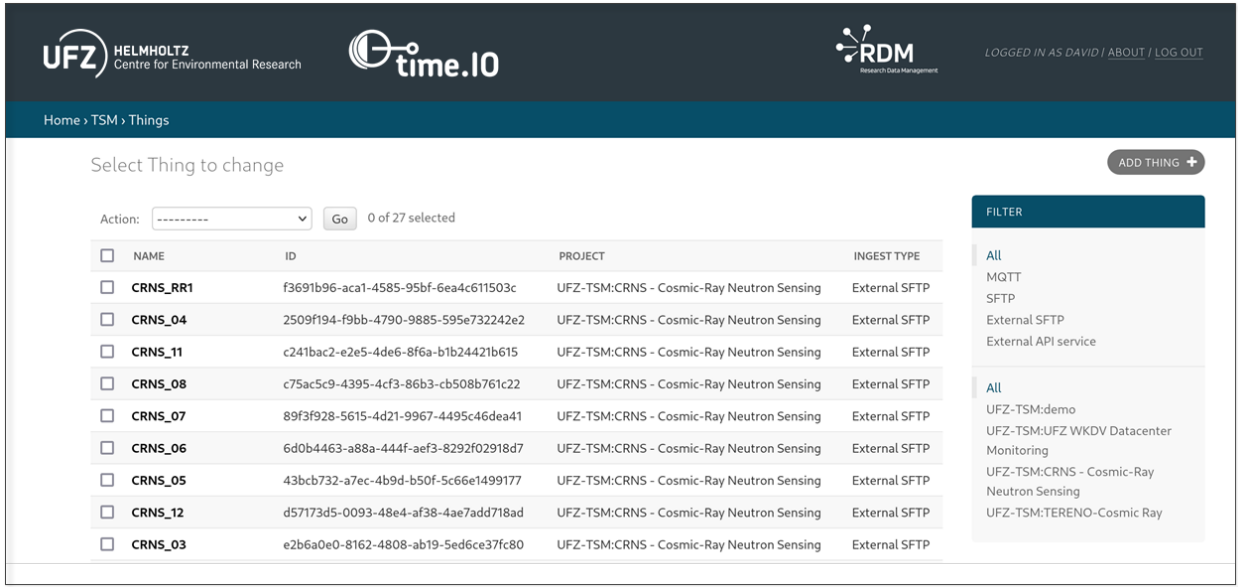

*Figure 3: Front-End time.IO. The time.IO is developed to handle the entire lifecycle of sensor based data streams.*

Its functionalities include:

- **Data transfer and storage**: Efficiently handles the transfer and storage of large volumes of time series data. time.IO provides transfer endpoints such as (S-)FTP servers or MQTT topics, but also supports the integration of existing data transfer infrastructures.
- **Data visualisation**: Uses Grafana[8] to provide real-time visualizations of time series data within automatically setup, preconfigured and shareable dashboards.
- **Quality control**: Integrates seamlessly with SaQC to ensure data quality and integrity.
- **Metadata management**: Uses the SMS for consistent and standardized metadata management.
- **Deployment**: Provides a container-based deployment model for easy integration and scalability within existing IT infrastructures.

For more details, visit: Schäfer et al. 2023 and/or https://codebase.helmholtz.cloud/ufz-tsm

---

[8] https://grafana.com/

## 2.3 System for Automated Quality Control (SaQC)

Completing the ecosystem, SaQC automates the quality control of time series data, improving traceability and reproducibility. It supports detailed data analysis based on a catalogue of state-of-the-art time series analysis, data processing, and annotation of data using predefined or custom quality schemes. The meticulous collection of metadata throughout all operations ensures that resulting data meets quality standards such as traceability and reproducibility. This automated quality control is essential to maintain the integrity of data used in environmental research (Schmidt et al. 2023).

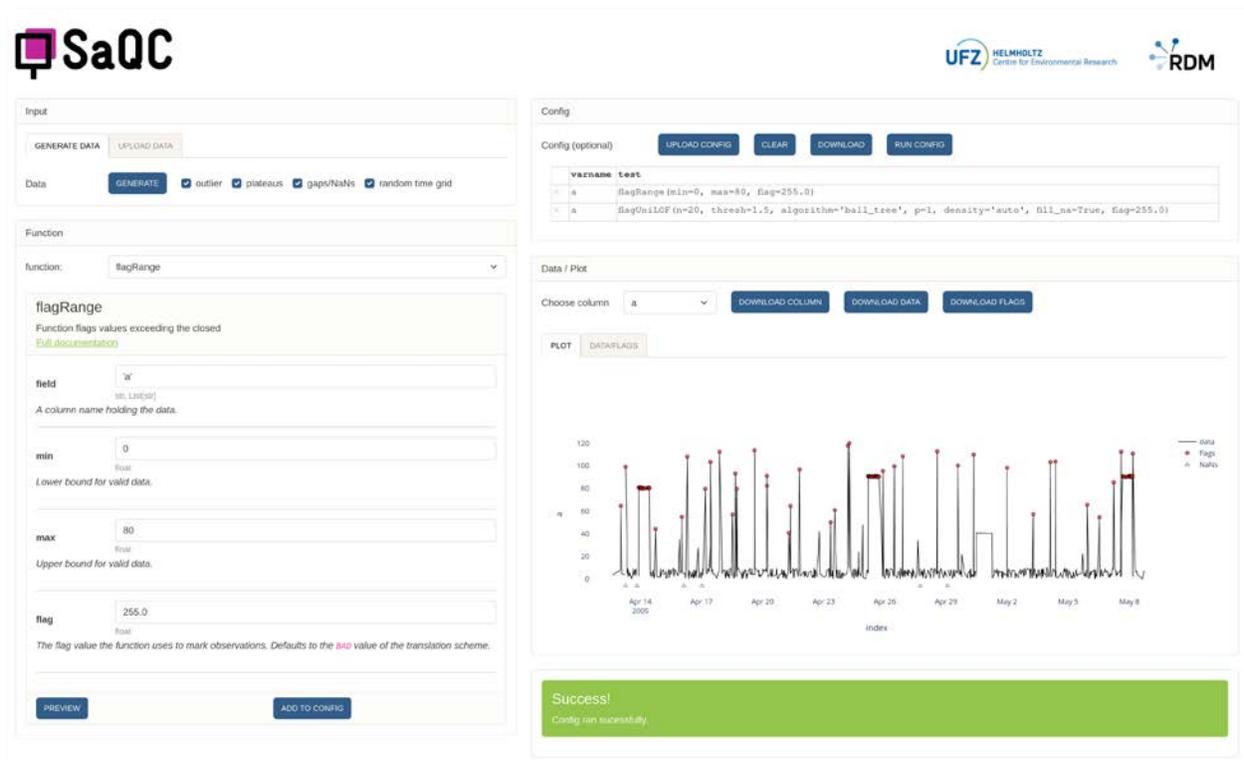

*Figure 4: Front-End SAQC (GUI): Users can add tests for their data in the panel on the left and see the resulting flags on the right side. These configurations can then be used for automated testing, allowing for continuous monitoring and quality assurance of the data.*

SaQC is designed to automate the quality control of time series data, improving traceability and reproducibility. Key features include:

- **Data Analysis:** Provides state of the art algorithms for detailed analysis of time series data.
- **Data Processing:** Exposes a large set of data processing features
- **Data Annotation**: Supports the annotation of time series data using predefined or custom quality schemes.
- **End-to-end metadata enrichment**: Enrich metadata from initial data collection to final use.
- **User interfaces**: Provides flexibility through a Python API, text-based configuration, and a web application.

For more details, visit: Schmidt et al. 2023, Schäfer et al. 2024 and/or https://git.ufz.de/rdm-software/saqc

## 3. Illustrative Example

The presented toolchain encompasses every stage of the typical sensor data lifecycle through user-friendly frontends, which include (i) metadata registration during sensor deployment, (ii) the setup of data transmission, including the automatic generation of endpoints, accounts, and credentials, (iii) real-time monitoring and provisioning of incoming data streams, and (iv) the configuration and parameterization of quality control pipelines. This self-service approach is designed specifically for data producers, data managers, and technicians, abstracting the technical complexities of underlying technologies such as data transfer and storage (see Figure 5).

**Cosmic Ray Neutron Sensing (CRNS)**

CRNS - Cosmic Ray Neutron Sensing is a method for determining average soil moisture (non-contact technology) for areas of about 5-15 hectares (Köhli et al. 2015). This novel approach to soil moisture measurement is one of our reference use cases for time.IO and well established in the TERENO community.

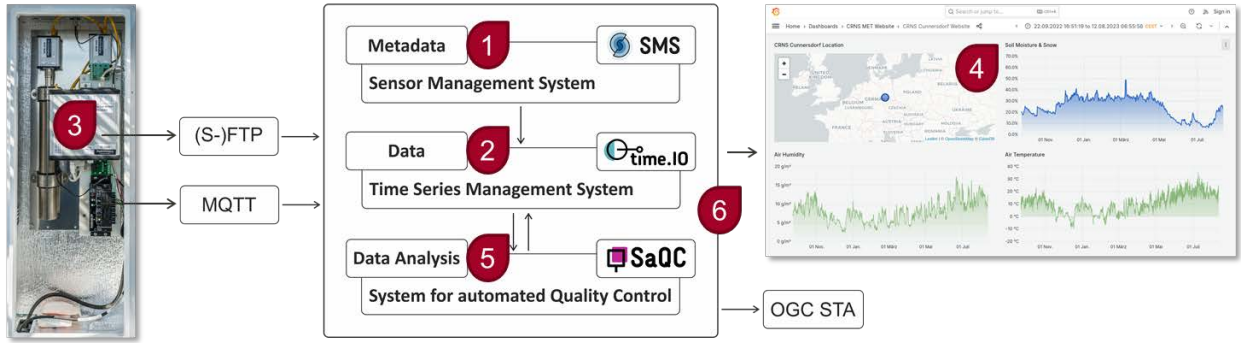

1. Register metadata with **SMS**
2. Setup Thing with **time.IO**
3. Configure logger and start data transfer
4. Monitor datastreams (**time.IO/Grafana**)
5. Setup quality control and attach custom processing steps (**SaQC**)
6. Share dashboards and data (**time.IO/Grafana/OGC STA**)

*Figure 5: The toolchain for implementing a CRNS sensor into the digital ecosystem and the necessary workflow steps.*

A typical usage pattern begins with registering the CRNS sensor's metadata in the Sensor Management System, where detailed information about the device—such as type, manufacturer, model, and the quantities it measures, including air temperature, air pressure, and neutron counts—is provided. For a real-world example, refer to the following link: https://web.app.ufz.de/sms/devices/443/basic. Next, a 'Thing' is created in the time.IO frontend. The term 'Thing' is derived from the OGC SensorThings API data model and is best described as a data transmission unit. In practice, a Thing often corresponds directly to a data logger or sensor. The user is required to input information such as hostnames for (S-)FTP servers or MQTT brokers, as well as account settings and credentials. Once the data transmission unit is operational, data monitoring can be conducted through Grafana. All incoming data becomes immediately visible in automatically set up and pre-configured dashboards. Subsequently, SaQC can be configured within the time.IO frontend to apply quality control and data processing directly within the data stream. Finally, data and metadata are provided both through the built-in Grafana dashboard for human users and via the OGC standard SensorThings API for machine-to-machine data exchange.

## 4. Impact

The digital ecosystem for managing time series data has significantly advanced research and practical applications in environmental system science. Below, we outline its impact across several key areas.

**New research opportunities and enhancing existing research**

The integration of SMS, time.IO, and SaQC within this digital ecosystem has opened up new research opportunities, particularly in real-time applications and data visualization. Automated, real-time quality control and immediate data viewers allow researchers to monitor and respond to environmental changes as they occur. This capability supports the study of dynamic events and enhances iterative research designs, enabling precise exploration of complex environmental interactions. The system's standardized framework also facilitates cross-disciplinary research by seamlessly integrating diverse sensor data, broadening the scope of environmental studies.

**Changes in Daily Practice**

Daily practices have evolved with the system's deployment. Automated workflows reduce manual data management, allowing researchers to focus on analysis rather than logistics. User-friendly interfaces make advanced data management accessible, boosting productivity and enabling more efficient handling of large datasets.

**Adoption and use of a transferable, standardized digital ecosystem**

The system's modular and cloud-ready architecture ensures its broad applicability and transferability across diverse research environments. By adopting standardized methods and protocols, the ecosystem enables consistent and reliable data management practices, making it possible to deploy identical infrastructures across different institutions. This standardization not only facilitates collaboration and data sharing but also enhances the comparability of research outcomes, allowing for more cohesive and integrated environmental studies globally. The system's versatility ensures that it can be effectively utilized in various scales of research, from localized projects to large-scale monitoring networks, driving innovation and efficiency in environmental data management.

## 5. Conclusions

The integration of the Sensor Management System (SMS), time.IO for storage, transfer, and real-time visualization, and the System for Automated Quality Control (SaQC) represents a significant advancement in the field of environmental data management. This digital ecosystem not only ensures the comprehensive management of time series data but also enhances data integrity, accessibility, and usability. By combining data acquisition, temporal alignment, and real-time quality control, the system supports robust environmental research and informed policy-making.

The modular, cloud-ready architecture and use of standardized protocols make the ecosystem highly adaptable and transferable across various research environments. This flexibility allows for consistent data management practices across institutions, fostering collaboration and enabling the comparability of research outcomes. The ecosystem's design ensures it can be deployed in both small-scale projects and large-scale monitoring networks, making it a versatile tool for advancing environmental science.

In summary, this digital ecosystem offers a powerful, standardized solution for managing environmental sensor data, supporting the pursuit of new research questions and improving the efficiency and reliability of existing studies. Its widespread adoption will likely drive further innovation in environmental monitoring and data management, benefiting a broad range of stakeholders in both scientific and practical contexts.

**Acknowledgements**

We thank the Helmholtz Association and the Federal Ministry of Education and Research (BMBF) for supporting the DataHub initiative of the Research Field Earth and Environment. The DataHub enables an overarching and comprehensive research data management, according to the FAIR principles, for all topics of the programme Changing Earth – Sustaining our Future.

The publication is a contribution in the context of the work of the NFDI4Earth – the consortium for the Earth System Sciences within the German National Research Data Infrastructure (NFDI) e.

V.. The NFDI is financed by the Federal Republic of Germany and its 16 federal states, and the NFDI4Earth is funded by the Deutsche Forschungsgemeinschaft (DFG, German Research Foundation) – project number: 460036893. The authors would like to express their gratitude for the funding and support.

This research has been supported by the European Commission, Horizon 2020 (H2020 Project eLTER PPP, grant no. 871126, eLTER PLUS, grant no. 871128 and eLTER EnRich, grant no. 101131751) in the context of the development of the eLTER cyberinfrastructure.

**References**

1. Stocker, T. F., et al. (2014): Climate Change 2013: The Physical Science Basis. Contribution of Working Group I to the Fifth Assessment Report of the Intergovernmental Panel on Climate Change. Cambridge University Press. https://doi.org/10.1017/CBO9781107415324

2. Loescher, H.W., Kelly, E.F. and Lea, R. (2017): National ecological observatory network: Beginnings, programmatic and scientific challenges, and ecological forecasting. Terrestrial Ecosystem Research Infrastructures. CRC Press, pp. 27–52.

3. Mollenhauer, H., Kasner, M., Haase, P., Peterseil, J., Wohner, C., Frenzel, M., Mirtl, M., Schima, R., Bumberger, J. and Zacharias, S. (2018): Long-term environmental monitoring infrastructures in europe: observations, measurements, scales, and socio-ecological representativeness. Sci. Total Environ. 624, 968–978. http://dx.doi.org/10.1016/j.scitotenv.2017.12.095

4. Ohnemus, T., Zacharias, S., Dirnböck, T., Bäck, J., Brack, W., Forsius, M., Mallast, U., Nikolaidis, N. P., Peterseil, J., Piscart, C., Pando, F., Terán, C. P. and Mirtl, M. (2024): The eLTER research infrastructure: Current design and coverage of environmental and socio-ecological gradients. Environmental and Sustainability Indicators 23, 100456. https://doi.org/10.1016/j.indic.2024.100456

5. Zacharias, S., Bogena, H., Samaniego, L., Mauder, M., Fuß, R., Pütz, T., Frenzel, M., Schwank, M., Baessler, C., Butterbach-Bahl, K., Bens, O., Borg, E., Brauer, A., Dietrich, P., Hajnsek, I., Helle, G., Kiese, R., Kunstmann, H., Klotz, S., Munch, J.C., Papen, H., Priesack, E., Schmid, H.P., Steinbrecher, R., Rosenbaum, U., Teutsch, G. and Vereecken, H. (2011): A Network of Terrestrial Environmental Observatories in Germany. Vadose Zone Journal, 10: 955-973. https://doi.org/10.2136/vzj2010.0139

6. Zacharias, S., Loescher, H.W., Bogena, H., Kiese, R., Schrön, M., Attinger, S., Blume, T., Borchardt, D., Borg, E., Bumberger, J., Chwala, C., Dietrich, P., Fersch, B., Frenzel, M., Gaillardet, J., Groh, J., Hajnsek, I., Itzerott, S., Kunkel, R., Kunstmann, H., Kunz, M., Liebner, S., Mirtl, M., Montzka, C., Musolff, A., Pütz, T., Rebmann, C., Rinke, K., Rode, M., Sachs, T., Samaniego, L., Schmid, H.P., Vogel, H.-J., Weber, U., Wollschläger, U. and Vereecken, H. (2024): Fifteen years of integrated Terrestrial Environmental Observatories (TERENO) in Germany: functions, services, and lessons learned, Earth Future 12 (6), https://doi.org/10.1029/2024EF004510

7. Weber, U., Attinger, S., Baschek, B., Boike, J., Borchardt, D., Brix, H., Brüggemann, N., Bussmann, I., Dietrich, P., Fischer, P., Greinert, J., Hajnsek, I., Kamjunke, N., Kerschke, D., Kiendler-Scharr, A., Körtzinger, A., Kottmeier, C., Merz, B., Merz, R., Riese, M., Schloter, M., Schmid, H., Schnitzler, J., Sachs, T., Schütze, C., Tillmann, R., Vereecken, H., Wieser, A. and Teutsch, G. (2022): MOSES: A Novel Observation System to Monitor Dynamic Events across Earth Compartments. Bulletin of the American Meteorological Society, 103(2), E339-E348. https://doi.org/10.1175/BAMS-D-20-0158.1

8. Wilkinson, M. D., et al. (2016): The FAIR Guiding Principles for scientific data management and stewardship. Scientific Data, 3, 160018. https://doi.org/10.1038/sdata.2016.18.

9. Mons, B., et al. (2017): Cloudy, increasingly FAIR; revisiting the FAIR Data guiding principles for the European Open Science Cloud. Information Services & Use, 37(1), 49-56. https://doi.org/10.3233/ISU-170824.

10. Koedel, U., Schuetze, C., Fischer, P., Bussmann, I., Sauer, P. K., Nixdorf, E., Kalbacher, T., Wichert, V., Rechid, D., Bouwer, L. M. and Dietrich, P. (2022): Challenges in the Evaluation of Observational Data Trustworthiness From a Data Producers Viewpoint (FAIR+). Front. Environ. Sci. 9:772666. https://doi.org/10.3389/fenvs.2021.772666

11. Selsam, P., Bumberger, J., Wellmann, T., Pause, M., Gey, R., Borg, E. and Lausch, A. (2024): Ecosystem Integrity Remote Sensing—Modelling and Service Tool—ESIS/Imalys. Remote Sensing, 16, 1139. https://doi.org/10.3390/rs16071139

12. Brinckmann, N., Alhaj Taha, K., Kuhnert, T., Abbrent, M., Becker, W., Bohring, H., Breier, J., Bumberger, J., Ecker, D., Eder, T., Gransee, F., Hanisch, M., Lorenz, C., Moorthy, R., Nendel, L. J., Pongratz, E., Remmler, P., Rosin, V., Schaeffer, M., Schaldach, M., Schäfer, D., Sielaff, D. and Ziegner, N. (2024): Sensor Management System - SMS (1.16.1). Zenodo. https://zenodo.org/doi/10.5281/zenodo.13329925

13. Schäfer, D., Abbrent, M., Gransee, F., Kuhnert, T., Hemmen, J., Nendel, L., Palm, B., Schaldach, M., Schulz, C., Schnicke, T. and Bumberger, J. (2023): timeIO - A fully integrated and comprehensive timeseries management system (0.1). Zenodo. https://zenodo.org/doi/10.5281/zenodo.8354839

14. Schmidt, L., Schäfer, D., Geller, J., Lünenschloss, P., Palm, B., Rinke, K., Rebmann, C., Rode, M. and Bumberger, J. (2023): System for automated Quality Control (SaQC) to enable traceable and reproducible data streams in environmental science. Environmental Modelling & Software, 105809. https://doi.org/10.1016/j.envsoft.2023.105809.

15. Schäfer, D., Palm, B., Lünenschloß, P., Schmidt, L., Schnicke, T. and Bumberger, J. (2024): System for automated Quality Control - SaQC (v2.6.0). Zenodo. https://zenodo.org/doi/10.5281/zenodo.5888547

16. Köhli, M., Schrön, M., Zreda, M., Schmidt, U., Dietrich, P. and Zacharias, S. (2015): Footprint characteristics revised for field-scale soil moisture monitoring with cosmic-ray neutrons. Water Resources Research, 51(7), 5772-5790. https://doi.org/10.1002/2015WR017169